\documentclass[aps,twocolumn,showpacs]{revtex4}
\usepackage{amsmath}

\usepackage{graphicx}
\usepackage{dcolumn}
\usepackage{bm}
\usepackage{color, soul}
\usepackage[utf8]{inputenc}
\usepackage[T1]{fontenc}
\usepackage{mathptmx}
\usepackage{multirow}
\usepackage{makecell}
\usepackage{setspace}
\usepackage{footnote}
\usepackage{hyperref}
\hypersetup{hypertex=true,
            colorlinks=true,
            linkcolor=blue,
            anchorcolor=blue,
            citecolor=blue}

\begin{document}
\title{Breathing solitons induced by collision in dipolar Bose-Einstein condensates}

\author{Peng Gao$^{1,2}$}
\author{Xin Li$^{1,2}$}
\author{Zhan-Ying Yang$^{1,2,3}$}\email{zyyang@nwu.edu.cn}
\author{Wen-Li Yang$^{1,2,3,4}$}
\author{Su Yi$^{5,6}$}

\address{$^1$School of Physics, Northwest University, Xi'an 710127, China}
\address{$^2$Shaanxi Key Laboratory for Theoretical Physics Frontiers, Xi'an 710127, China}
\address{$^3$Peng Huanwu Center for Fundamental Theory, Xi'an 710127, China}
\address{$^4$Institute of Modern Physics, Northwest University, Xi'an 710127, China}
\address{$^5$CAS Key Laboratory of Theoretical Physics, Institute of Theoretical Physics, Chinese Academy of Sciences, Beijing 100190, China}
\address{$^6$School of Physics \& CAS Center for Excellence in Topological Quantum Computation, University of Chinese Academy of Sciences, Beijing 100190, China}

\begin{abstract}
We numerically study the breathing dynamics induced by collision between bright solitons in the one-dimensional Bose-Einstein condensates with strong dipole-dipole interaction.
This breathing phenomenon is closely related to the after-collision short-lived attraction of solitons induced by the dipolar effect.
The initial phase difference of solitons leads to the asymmetric dynamics after collision, which is manifested on their different breathing amplitude, breathing frequency, and atom number.
We clarify that the asymmetry of breathing frequency is directly induced by the asymmetric atom number, rather than initial phase difference.
Moreover, the collision between breathing solitons can produce new after-two-collision breathing solitons, whose breathing amplitude can be adjusted and reach the maximum (or minimum) when the peak-peak (or dip-dip) collision happens.
\end{abstract}

\maketitle

\section{Introduction}
The condensates of bosonic gases with large magnet dipole moments have been created \cite{Griesmaier-2005,Beaufils-2008,Lu-2011,Tang-2014,Aikawa-2012,Stuhler-2005}.
Contrary to the contact interaction between atoms, the dipole-dipole interaction is long-range and anisotropic, so it brings many novel influences on characters of condensates and the localized structures including solitons \cite{Dauxois-2002,Giovanazzi-2002,Sinha-2007,Kevrekidis-2007,Koch-2008,Lahaye-2008,Lahaye-2009,Kadau-2016,
Ferrier-Barbut-2016}.
In the dipolar Bose-Einstein condensates (BECs) described by the Gross-Pitaevskii equation, many efforts have been made to study these localized structures.
Two-dimensional stable bright and vortex solitons were generated in the cases without external traps, which benefited from the long-range feature of dipolar effect \cite{Pedri-2005,Lashkin-2007}; the formation of anisotropic bright and vortex solitons was closely related to its anisotropic feature \cite{Raghunandan-2015,Tikhonenkov-2008,Yi-2006}.
Some new localized waves have been also found in one-dimensional dipolar BECs, like soliton molecules and the dark soliton with ripples \cite{Baizakov-2015,Turmanov-2015,Pawowski-2015,Edmonds-2016}.
Meanwhile, the interaction between dipolar solitons exhibited various dynamical behaviors.
Two solitons may merge into a breathing wave when the dipolar interaction was considered \cite{Cuevas-2009,Pedri-2005,Edmonds-2017}; they could rebound from each other in some other cases \cite{Cuevas-2009,Eichler-2012}.
When the initial distance between two solitons is small, two out-of-phase solitons could form a bound state with a stable oscillating frequency \cite{Bland-2015,Baizakov-2015,Pawowski-2015}.
These phenomena indicate that the interaction between solitons becomes inelastic under the dipolar effect.

Recently, breathing solitons also draw the attention of scientists due to their character of stable breathing.
Their dynamical features and generation mechanism have been discussed in real physical systems\cite{Peng-2019,Bao-2018,Lucas-2017,Bao-2016,Yu-2017,Bao-2015,Matsko-2012}.
As we know, the breathing dynamics of solitons cannot be induced after bright solitons collide in one-dimensional non-dipolar BECs.
However, the breathing phenomenon appears after the collision between dipolar solitons.
It provides a new way to generate and control breathing solitons in dipolar BECs so is worthy of more attention.

In this paper, we numerically study the breathing character of bright solitons induced by their collision in the one-dimensional BECs with strong dipole-dipole interaction.
From two solitons with identical density distribution, they have different breathing amplitude, breathing frequency, and atom number after collision, under different initial phase difference of them.
We demonstrate that the asymmetry of breathing frequency is directly related to the asymmetric distribution of after-collision atom number, rather than the initial phase difference.
When we make the peaks of two breathing solitons collide, their breathing amplitude gets larger after the collision, and conversely a dip-dip collision generates solitons with smaller breathing amplitude.
Thus, one can strengthen or weaken the breathing character by the collision between breathing solitons.
Importantly, with the breathing amplitude increasing, the derived solitons have lower values of velocity and mean kinetic energy.
It manifests that there is a short-lived attraction of solitons after collision, which is the major cause of breathing phenomenon.

\section{Model of the dipolar BECs}

The dynamics of dipolar BECs can be discribed by the dipolar Gross-Pitaevskii model \cite{Kevrekidis-2007,Baranov-2008,Lahaye-2009},
\begin{eqnarray}
\begin{split}
\label{eq-model3d}
i\hbar\frac{\partial\Psi}{\partial t}=-\frac{\hbar^2}{2m}\nabla^2\Psi+V(\mathbf{r})\Psi+g|\Psi|^2\Psi+\Phi_{\rm{dd}}(\mathbf{r},t)\Psi,
\end{split}
\end{eqnarray}
where $\Psi(\mathbf{r},t)$ is the mean-field wave function of condensates. The strength of contact interaction is $g=4\pi\hbar^2a_s/m$ with the atom mass $m$ and the $s$-wave scattering length $a_s$.
The external potential $V(\mathbf{r})$ provides a trap where the cloud is confined, and it can be assumed as a harmonic form, $V(\mathbf{r})=m\omega_\perp^2 (y^2+z^2)/2$, where $\omega_\perp$ is the transverse trapping frequency.
Here, the axial trap is neglected.
The nonlocal dipolar potential is $$\Phi_{\rm{dd}}(\mathbf{r},t)=\int \frac{d^2(1-3\cos^2\theta_{rd})}{|\mathbf{r}-\mathbf{r}'|^3}|\Psi(\mathbf{r}',t)|^2 d\mathbf{r}',$$
where $d$ is the magnetic dipole moment and $\theta_{rd}$ is the angle between the vector joining interacting particles and the dipole direction.

In the transverse external trap $V(\mathbf{r})$, the condensates are compressed to be cigar-shaped and can be considered as quasi-one-dimensional dipolar BECs.
Thus, the 3D model (\ref{eq-model3d}) can be reduced into a 1D model.
To this end, an effective way is to assmue the wave function $\Psi(\mathbf{r},t)=\tilde{\psi}(x,t)\psi_\perp(y,z)$, where $\psi_\perp(y,z)=\exp[-(y^2+z^2)/2l_\perp^2]/(l_\perp\sqrt{\pi})$ and $l_\perp=\sqrt{\hbar/m\omega_\perp}$.
By the integration over transverse directions, the effective 1D potential of dipole-dipole interaction is obtained, $$\tilde{\Phi}_{\rm{dd}}(x,t)=-\frac{\alpha d^2}{2l_\perp^3}\int^{+\infty}_{-\infty} \tilde{R}(x-x')|\tilde{\psi}(x',t)|^2 dx',$$
where $\alpha=(1+3\cos2\theta)/4$ has the range from $-1/2$ to $1$, and $\theta$ is the angle between dipole and $x$ directions. 
The nonlocal response function has the form of $\tilde{R}(x)=(1+2u^2)\exp(u^2){\rm erfc}(|u|)-2{\pi}^{-1/2}|u|$, where $u=x/\sqrt{2}l_\perp$ and erfc is the complementary error function.
(The detailed calculation of 1D dipolar potential was illustrated in Ref. \cite{Deuretzbacher-2010}.)
Then, by the transformation
\begin{eqnarray}
\begin{split}
t=\frac{2}{\omega_\perp}T, \quad x=\sqrt{2}l_\perp X, \quad \tilde{\psi}=\frac{1}{2\sqrt{|a_{s0}|}}\psi, \nonumber
\end{split}
\end{eqnarray}
we can transform the 1D dipolar model into a dimensionless one.
For convenience, the labels $X$ and $T$ are severally replaced by $x$ and $t$, and the final dimensionless model becomes
\begin{eqnarray}
\begin{split}
\label{eq-model}
i{\psi}_{t}&+\frac{1}{2}{\psi}_{xx}+g_c|{\psi}|^2{\psi}\\
&+g_d{\psi}{\int_{-\infty}^{+\infty}R(x-x')|{\psi(x',t)}|^2dx'}=0,
\end{split}
\end{eqnarray}
where the responce function is $${R}(x)=(1+2x^2)\exp(x^2){\rm erfc}(|x|)-2{\pi}^{-1/2}|x|.$$
In this equation, $g_c=-{a_s}/{|a_{s0}|}$ and $g_d={\sqrt{\pi}\alpha a_d}/{2|a_{s0}|}$ scale the strength of contact and dipolar interactions, respectively, where $a_d=md^2/\hbar^2$ is the characteristic dipole length and $a_{s0}$ is the background value of scattering length.
The positive(or negative) values of $g_c$ and $g_d$ indicates the attractive (or repulsive) interaction.

In our numerical simulations, the dipolar condensates of $^{164}$Dy atoms are considered, which have strong dipole-dipole and weak contact interactions.
By applying an external magnetic field and Feshbach resonance technique, the condition $a_s\rightarrow 0$ can be satisfied for the case of strongly dipolar interaction.
We have $m=2.7\times 10^{-25}{\rm kg}$, $\omega_\perp=2\pi\times 250 {\rm Hz}$, and $a_{s0}=1.17\times 10^{-11}{\rm m}$, which leads to $l_\perp=1.25 {\rm \mu m}$.
The units of space and time coordinates are respectively $x_{\rm u}=\sqrt{2}l_\perp=1.77 {\rm \mu m}$ and $t_{\rm u}=2/\omega_\perp=1.27 {\rm ms}$.
The realistic atom number is $\tilde{N}=N\sqrt{2}l_\perp/4|a_{s0}|=37718N$, where $N$ is the dimensionless one.
The dipole moment is $d=10\mu_B\sqrt{\mu_0/4\pi}=0.02\sqrt{\hbar\omega_\perp l_\perp^3}$, where $\mu_B$ is the Bohr magneton and $\mu_0$ is the permeability of vacuum, so $a_d=5.29\times 10^{-10}{\rm m}$ \cite{Lu-2011}.
The dipolar strength $g_d$ can be changed by adjusting $\alpha$.
For examples, we can set $g_d=20$, $30$, and $40$ to consider the cases of $\alpha=1/4$, $3/8$, and $1/2$, respectively.

\begin{figure}[htbp]
	\centering
	\includegraphics[width=82mm]{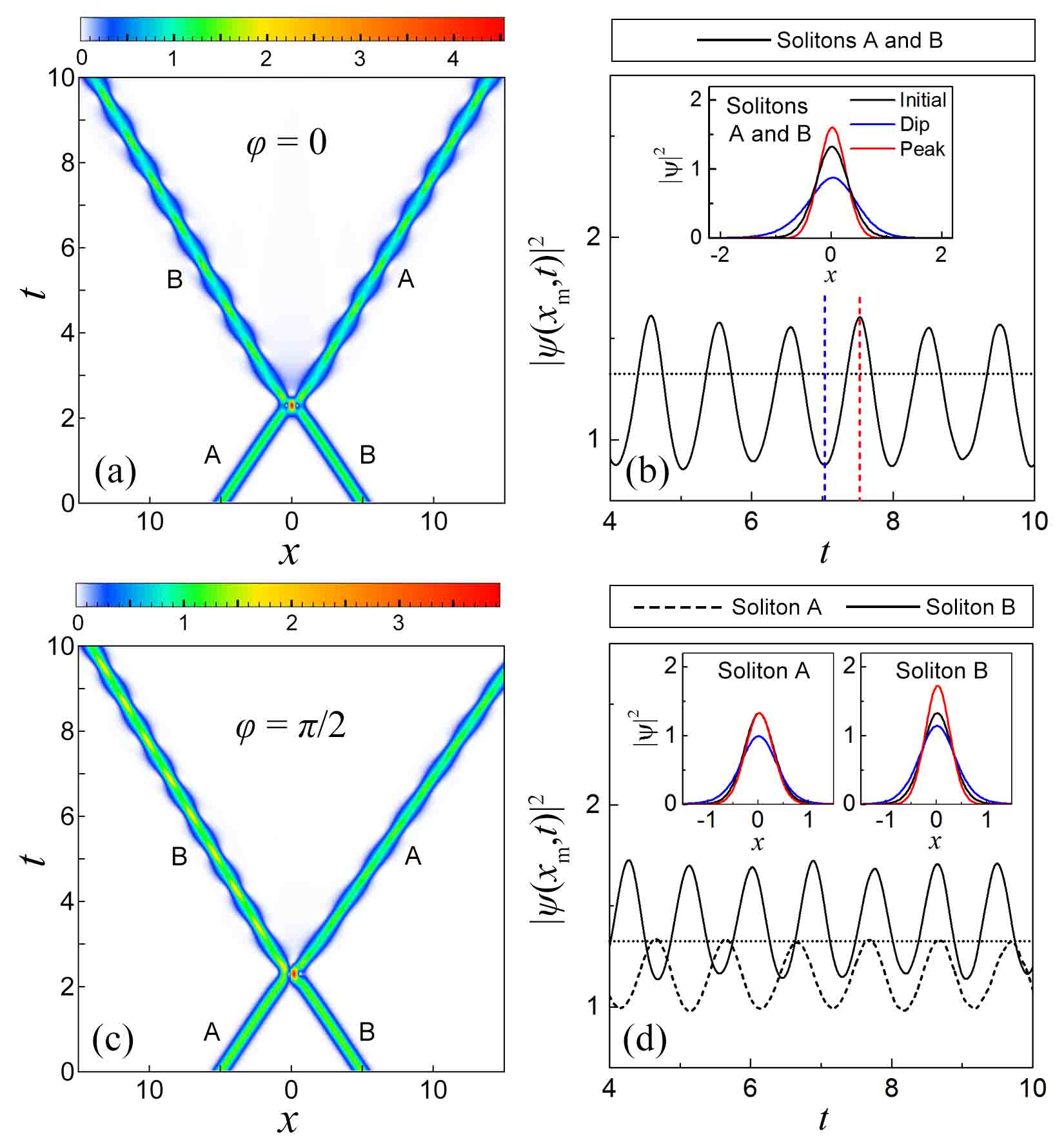}
	\caption{(Color online) (a) Density evolution plot of initial condition (\ref{eq-ini}) when $N=1$, $x_0=5$, $v=2$, $\varphi=0$, $g_c=0$, and $g_d=20$.
		(b) Its evolution of maximal density of two solitons. The black point line denotes the maximal density of initial soliton. The red and blue dashed lines denotes the time of peak and dip, respectively. The inset plot shows the density profiles of the initial soliton (black curve), the dip (blue curve) and peak (red curve) of breathing soliton.
		(c) Density evolution plot of initial condition (\ref{eq-ini}) when $\varphi$ is changed into $\pi/2$.
		(d) Its evolution of maximal density of Soliton A (dashed curve) and Soliton B (solid curve).
	}
	\label{pic-bs}
\end{figure}

\section{Breathing dynamics of solitons induced by collision}

\begin{figure*}[htpb]
	\centering
	\includegraphics[width=166mm]{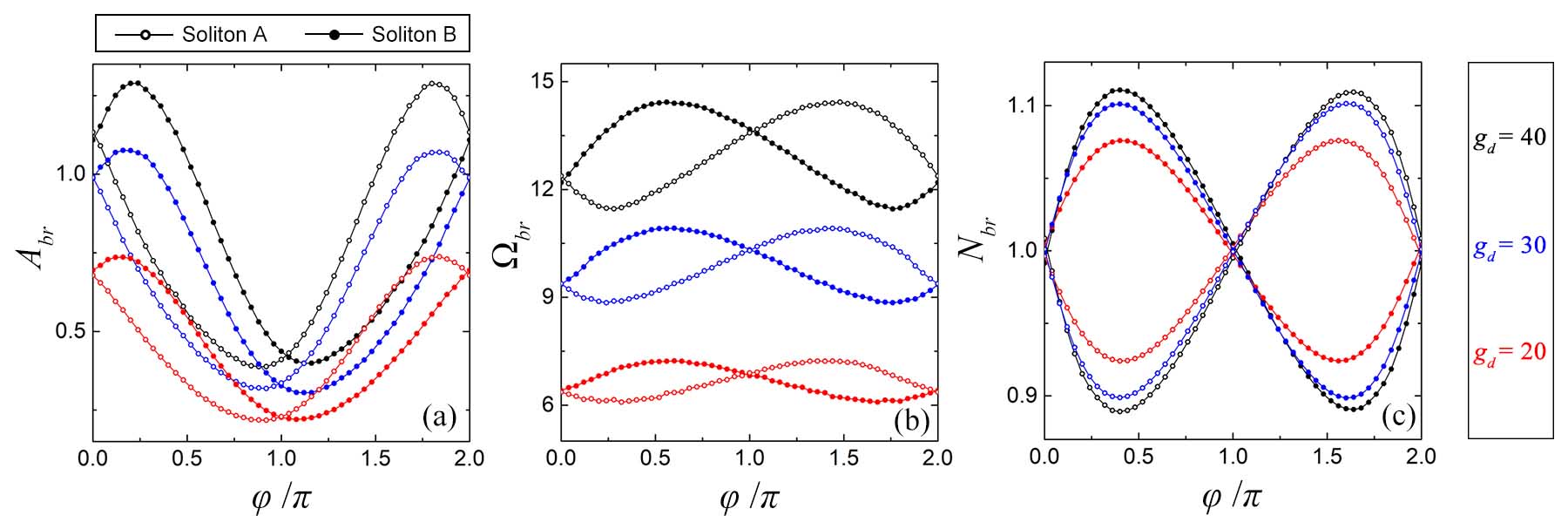}
	\caption{(Color online) Influence of phase difference $\varphi$ on the (a) breathing amplitude $A_{br}$, (b) breathing frequency $\Omega_{br}$, and  (c) after-collision atom number $N_{br}$ of solitons, when $g_d=20$ (red), $g_d=30$ (blue), and $g_d=40$ (black). The open and solid circles denote the results of Soliton A and B, respectively.}
	\label{pic-phi}
\end{figure*}

The bright solitons in the model (\ref{eq-model}) have been numerically studied, which contains the cases of dipolar and local interactions with identical or opposite types \cite{Cuevas-2009,Edmonds-2017}.
Here, we use the split-step Crank-Nicolson method and the imaginary-time propagation method \cite{Muruganandam-2003,Muruganandam-2009} to obtain the numerical profiles of bright solitons in this model.
They are denoted as $\psi_{0}(x)$, so the atom number of soliton is $N=\int_{-\infty}^{+\infty}|{\psi_0(x)}|^2dx$.
With the chemical potential $\mu$, the bright soliton solution in the model (\ref{eq-model}) can be written as $\psi_{bs}(x,t)=\psi_0(x)\exp(-i{\mu}t)$.
To study the collision between the solitons, we have the initial condition of two solitons with velocity and phase difference,
\begin{eqnarray}
\begin{split}
\label{eq-ini}
\psi(x,0)=\psi_0(x+x_0)e^{ivx}+\psi_0(x-x_0)e^{-ivx+i{\varphi}},
\end{split}
\end{eqnarray}
where $x_0>0$, $v\ge 0$, and $\varphi$ are the initial offset, velocity, and phase difference of solitons, respectively.
Here, we call the initial soliton on the left-hand side as Soliton A, while the one on the right-hand side as Soliton B.

We consider the cases of strongly dipolar interaction ($g_d=20$) and neglect the contact one ($g_c=0$).
The set of initial parameters is $N=1$, $x_0=5$, $v=2$, and $\varphi=0$, which corresponds to the collision between two in-phase solitons.
By the split-step Fourier method \cite{Yang-book}, its evolution plot of density $|\psi(x,t)|^2$ is shown in Fig. \ref{pic-bs} (a).
The two solitons become breathing with a stable frequency after they collide.
In the cases without collision, the breathing mode of dipolar solitons has been described by the variational approximation in three-dimensional BECs \cite{Yi-2001,Yi-2002}.
Here, the breathing phenomenon induced by soliton collision provides a new way to generate breathing solitons in dipolar BECs.
It is related to the dipolar effect, and it doesn't happen in the cases without dipolar effect ($g_d=0$), where the interaction between solitons are elastic.
Note that the solitons changes their positions on the space coordinate after the collision, namely, the after-collision Soliton A (or B) locates on the right-hand (or left-hand) side.
For the time $t$, we define the space coordinate where the soliton has the maximal density as $x_{\rm m}(t)$.
Thus, the maximal density of solitons can be written as $|\psi(x_{\rm m},t)|^2$.
Their evolutions of $|\psi(x_{\rm m},t)|^2$ are shown in Fig. \ref{pic-bs} (b). The density profiles of initial soliton and the breathing solitons at the time of peak and dip are shown in the inset.
The peak has a high and narrow profile while the dip has a short and wide profile, and the initial soliton has a profile in between.
Then, we change the phase difference into $\varphi=\pi/2$, which leads to an asymmetric collision.
Its evolution plots of density and maximal density of solitons are shown in Fig. \ref{pic-bs} (c) and (d), respectively.
The two solitons have different breathing dynamics after the collision.
The breathing amplitude of Soliton B is higher than the one of Soliton A, though both of them are smaller than the one induced by in-phase solitons in Figs. \ref{pic-bs} (a) and (b).

\begin{figure*}[htbp]
\centering
\includegraphics[width=160mm]{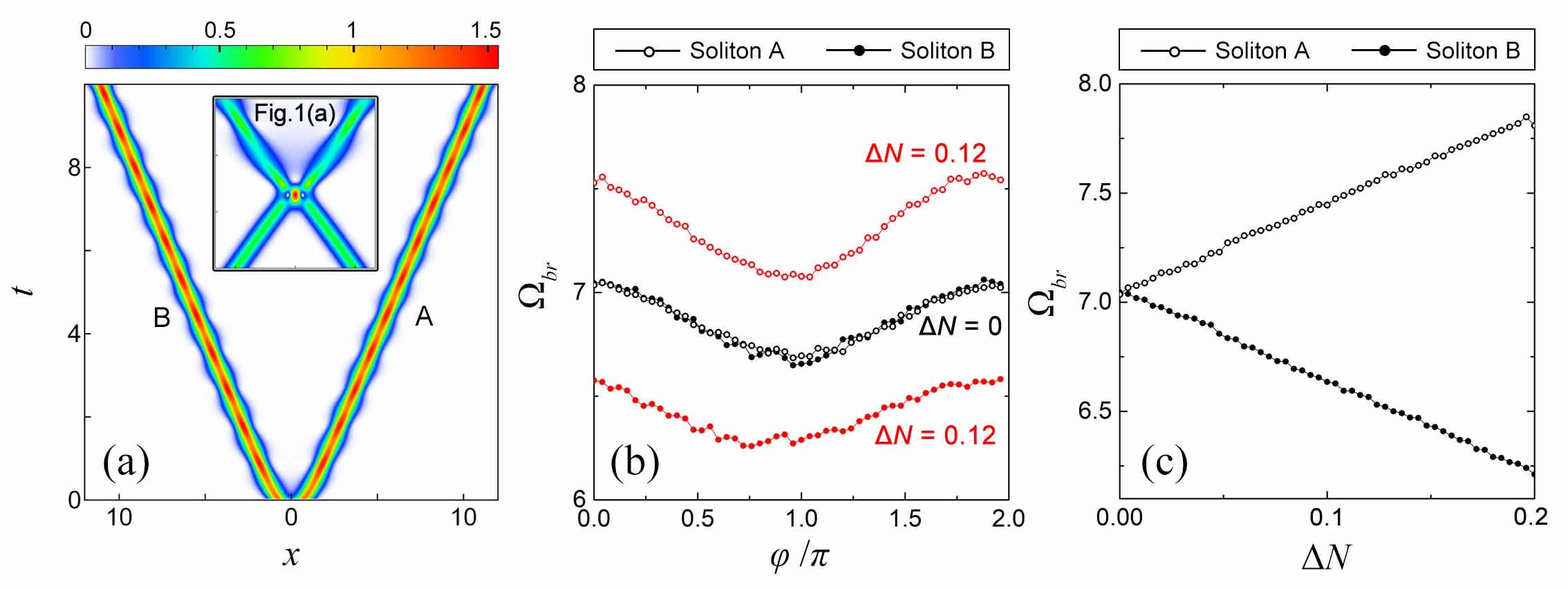}
\caption{(Color online) (a) Density evolution plot of initial condition (\ref{eq-ini}) when $N=1$, $x_0=0.7$, $v=-2$, and $\varphi=0$. It describes a process that two solitons move away from each other. The inset is a part of Fig. \ref{pic-bs} (a), describing the short-lived attracting process after collision. (b) Influence of phase difference $\varphi$ on the breathing frequency $\Omega_{br}$ of solitons, when $\Delta N=0$ (black) and $\Delta N=0.12$ (red). (c) Influence of atom number difference $\Delta N$ on the breathing frequency $\Omega_{br}$ when $\varphi=0$. The open and solid circle denote the results of Soliton A and B, respectively.}
\label{pic-dn}
\end{figure*}

From Fig. \ref{pic-bs}, one can find that the phase difference has an obvious influence on the breathing dynamics of solitons.
The breathing character of solitons can be depicted by its breathing amplitude $A_{br}$ and frequency $\Omega_{br}$.
We define the time when peak and dip appear as $t_{\rm peak}$ and $t_{\rm dip}$, respectively.
In the numerical simulation, the breathing period $T_{br}$ can be measured by the average value of several time differences of two adjacent peaks of $|\psi(x_{\rm m},t)|^2$, i.e., the average value of $t_{\rm peak}^{(n)}-t_{\rm peak}^{(n+1)}$.
Thus, the breathing frequency is $\Omega_{br}=2\pi/T_{br}$.
The breathing amplitude $A_{br}$ can be measured by the $|\psi(x_{\rm m},t)|^2$ difference at a pair of adjacent dip and peak, i.e., $A_{br}=|\psi(x_{\rm m},t_{\rm peak})|^2-|\psi(x_{\rm m},t_{\rm dip})|^2$.
The atom number of solitons after collision is denoted by $N_{br}$.
We illustrate the change of $A_{br}$, $\Omega_{br}$, and $N_{br}$ under the different phase difference $\varphi$ and dipolar strength $g_d$ in Figs. \ref{pic-phi} (a), (b), and (c), respectively.
When $\varphi\neq0$ and $\varphi\neq\pi$, the breathing characters of Soliton A and B are different after they collide.
When $0<\varphi<\pi$, both of $A_{br}$, $\Omega_{br}$, and $N_{br}$ of Soliton A are smaller than the ones of Soliton B; when $\pi<\varphi<2\pi$, the opposite results are obtained.
These results are obviously different from the cases without dipolar effect.
When $\varphi=0$ or $\varphi=\pi$, the two solitons have identical values of $A_{br}$, $\Omega_{br}$, and $N_{br}$.
Obvious decreases of $A_{br}$ happen when $\varphi$ is near $\pi$, which indicates that the breathing character is weakened in these cases.
Meanwhile, with the increase of $g_d$, both of $A_{br}$ and $\Omega_{br}$ increase, and the atom number difference $|N_{br}-1|$ also increases.
Note that the dependence of $N_{br}$ on $\varphi$ in Fig. \ref{pic-phi} (c) is quite similar to the one when two non-dipolar solitons collide on a narrow barrier \cite{Martin-2012,Helm-2012}.
Thus, the dipolar effect may have a barrier-like behavior for soliton collision.

By carefully observing the soliton collision in Fig. \ref{pic-bs} (a), there exists a short-lived attracting process between solitons after collision [see the inset of Fig. \ref{pic-dn} (a)].
The long-range attraction induced by dipolar effect leads to the widening of solitons and the subsequent breathing dynamics.
It indicates that the breathing solitons can be also generated from two solitons away from each other.
Its initial condition is still Eq. \ref{eq-ini}, and we set $v<0$ to make solitons move in the opposite directions.
When $N=1$, $x_0=0.7$, $v=-2$, and $\varphi=0$, the evolution plot is shown in Fig. \ref{pic-dn} (a).
The two solitons become breathing after they seperate.
Note that the after-collision atom number difference $\Delta N_{br}$ is identical with the initial one $\Delta N$ and is not influenced by the initial phase difference $\varphi$.
Thus, compared with soliton collisions, this way to generate breathing solitons allows us to severally study the influence of $\varphi$ and $\Delta N_{br}$, both of which can be set in the initial condition.
We set the initial atom number of two solitons are $1+\Delta N/2$ and $1-\Delta N/2$, namely, their atom number difference is $\Delta N$.
The influence of $\varphi$ on $\Omega_{br}$ when $\Delta N=0$ and $\Delta N=0.12$ is shown in Fig. \ref{pic-dn} (b).
The value of $N_{br}$ is measured at $t=15$.
Under different $\varphi$, when $\Delta N=0$, the breathing frequency $\Omega_{br}$ of two solitons is almost identical;
when $\Delta N=0.12$, their breathing frequency $\Omega_{br}$ is different and has an almost constant difference $\Delta\Omega_{br}$ between solitons.
Thus, the direct inducement of asymmetric breathing frequency between Solitons A and B is their atom number difference, rather than their phase difference.
Also, when $\varphi=0$, the dependence of $\Omega_{br}$ on $\Delta N$ is shown in Fig. \ref{pic-dn} (c).
The difference of initial atom number $\Delta N$ has a directly proportional influence on the difference of breathing frequency $\Omega_{br}$.
These results indicate that there is a positive correlation between $\Delta \Omega_{br}$ and $\Delta N$ (or $\Delta N_{br}$), but $\Delta \Omega_{br}$ is not influenced by $\varphi$.
Besides, it is found that the breathing amplitude $A_{br}$ is impacted by both of $\varphi$ and $\Delta N$, so their results are not shown in this paper.
Now, we look back on the process that two solitons collide and become breathing ones.
The change of $\varphi$ leads to the asymmetry of after-collision atom number $N_{br}$, and then the asymmetry of $N_{br}$ leads to the asymmetry of breathing frequency $\Omega_{br}$.
However, why the phase difference can cause the asymmetry of $N_{br}$ is still an open problem for the collision of dipolar solitons.

\begin{figure}[htbp]
	\centering
	\includegraphics[width=86mm]{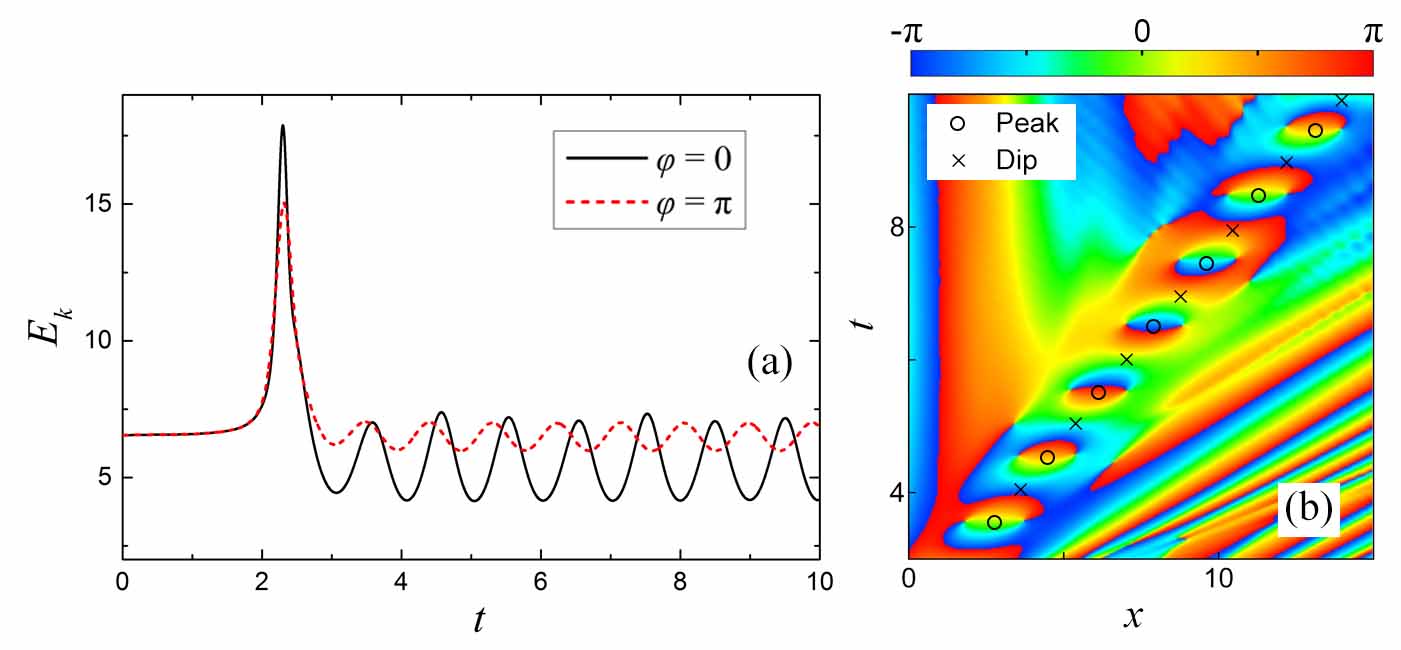}
	\caption{(Color online) (a) Kinetic energy evolution of soliton collision when $\varphi=0$ (black solid curve) and $\varphi=\pi$ (red dashed curve).
	(b) Phase evolution plot of Soliton A in Fig. \ref{pic-bs} (a), where the contribution of momentum is neglected. The circle and cross signs denote the positions of peaks and dips, respectively.
	The parameters are $N=1$, $x_0=5$, $v=2$, $g_c=0$, and $g_d=20$.
	}
	\label{pic-phase}
\end{figure}

We recall that the velocity of breathing soliton is smaller than the initial soliton's (namely $v_{br}<v$).
It may be related to the short-lived attracting process mentioned above.
The attraction makes the solitons wider and makes the soliton centers closer to each other than non-dipolar cases, so they have the smaller velocity than the initial solitons.
The widened solitons then become narrow and finally form the breathing ones.
This process can be also understood by the evolution of kinetic energy as shown in Fig. \ref{pic-phase} (b).
The kinetic energy of breathing solitons is fluctuating and has a lower mean value than the initial solitons', which agrees well with the comparison of their velocity.
Besides, it is known from Fig. \ref{pic-phi} (a) that the two derived breathing solitons when $\varphi=0$ have larger breathing amplitude than the ones when $\varphi=\pi$.
The evolution of their kinetic energy is compared in Fig. \ref{pic-phase} (b).
The kinetic energy when $\varphi=\pi$ has smaller fluctuation but nearly the same maximal amplitude than the one when $\varphi=0$.
Thus, the former has a higher mean value than the latter, which indicates that the out-of-phase breathing solitons have a larger velocity than the in-phase one.
We speculate that, with the breathing amplitude increasing, the breathing solitons have smaller velocity.

Finally, to analyze the fluxion of atoms in breathing solitons, we also study the evolution of their phase $\phi$.
The atomic velocity is the derivative of phase with respect to space coordinate, i.e., $v_{\rm atom}=\partial \phi/\partial x$.
The value of phase is calculated by $\phi= {\rm Arg}[\psi \exp({-iv_{br}x})]$ to neglect the contribution of momentum, where $v_{br}$ is the velocity of breathing soliton.
The phase evolution of Soliton A in Fig. \ref{pic-phi} (a) is shown in Fig. \ref{pic-phase} (a), and we have $v_{br}=1.74$ here.
When $t=t_{\rm peak}$ or $t=t_{\rm dip}$, the phase distribution of soliton is almost constant, which indicates the motionless atoms.
When $t=t_{\rm peak}+T_{br}/4$, the phase has smaller value at the soliton center and larger value at two sides, which indicates the expansion of soliton.
When $t=t_{\rm dip}+T_{br}/4$, the phase has larger center value and smaller two-side value, which indicates the compression of soliton.
Thus, in the breathing process, the atoms of soliton repeatedly gather and then disperse with a stable frequency.
Actually, the colliding process of two solitons is similar to the movement of a soliton away from the external trap formed by another soliton.
It indicates that the breathing phenomenon also appears when the soliton moves in a trap, which is observed in our numerical simulations.
In our opinion, the different parts of soliton distributing in trap have different acceleration.
Under the effect of acceleration difference, the soliton will become wider or narrower with the time evolution, and finally its breathing behavior manifests.

\begin{figure}[htbp]
	\centering
	\includegraphics[width=86mm]{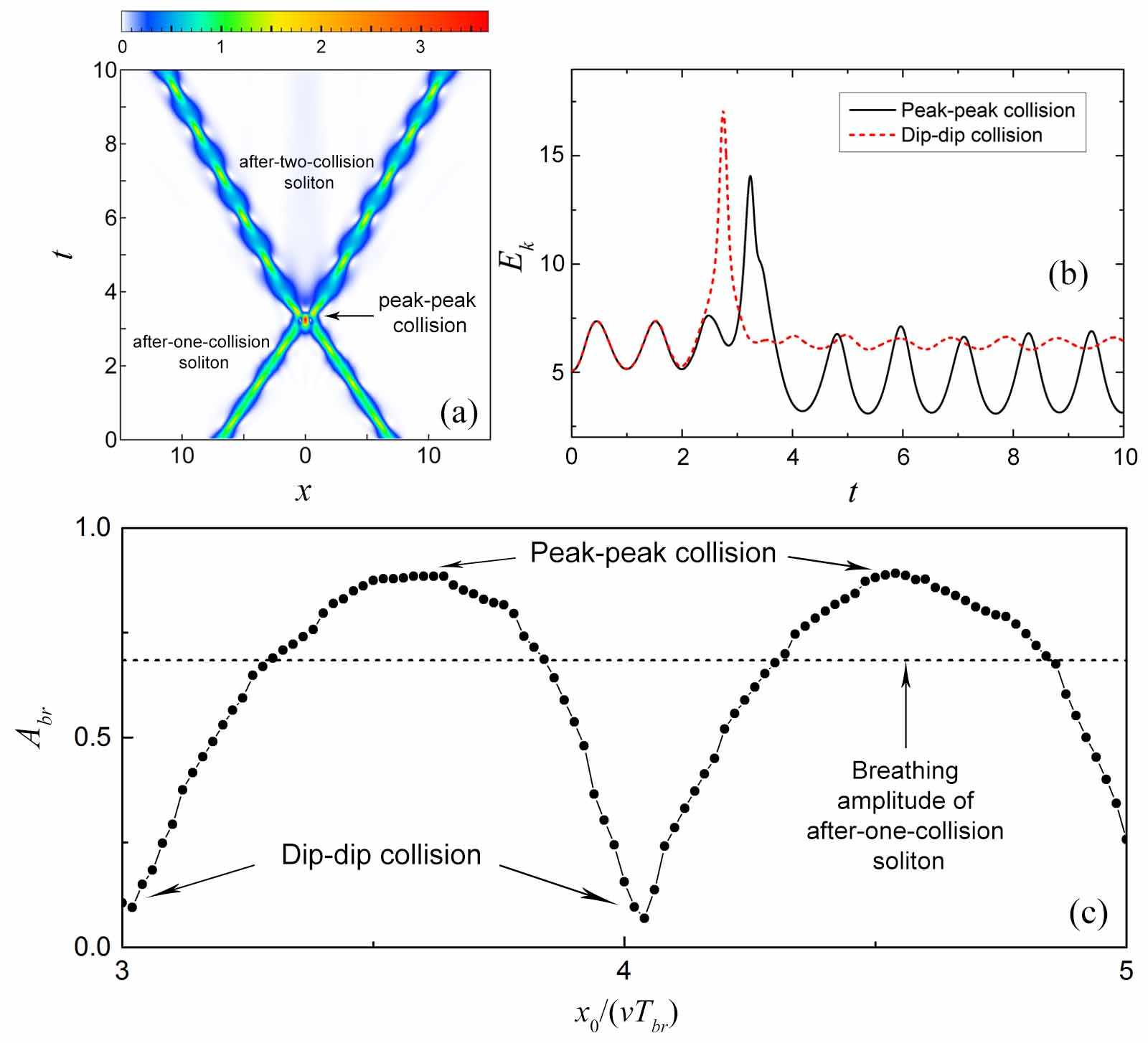}
	\caption{(Color online) (a) Density evolution plot for the interaction between two after-one-collision solitons. Two solitons with a larger breathing amplitude appear after a peak-peak collision happens.
	(b) Kinetic energy evolution of soliton collision in the cases of peak-peak (black solid curve) and dip-dip (red dashed curve) collisions.
	(c) Influence of initial offset on the breathing amplitude of after-two-collision solitons. The dashed black line denotes the breathing amplitude of after-one-collision solitons.}
	\label{pic-two}
\end{figure}

\section{Interaction between breathing solitons}

Since the collision between dipolar solitons can induce their breathing character, we pay attention to the influence of collision between breathing solitons on their breathing dynamics.
The initial condition is changed into
\begin{eqnarray}
\begin{split}
\label{eq-inibr}
\psi(x,0)=\psi_{br}(x+x_0)e^{ivx}+\psi_{br}(x-x_0)e^{-ivx},
\end{split}
\end{eqnarray}
where $\psi_{br}(x)$ is the dip profile of breathing solitons in Fig. \ref{pic-bs} (a), and the velocity $v$ is still set as $2$.
Note that the dip profile is filtered by a super-Gaussian function to weaken the radiation waves outside the solitons.
First, we focus on the cases that the collision happens at the peaks of two breathing solitons.
To this end, we have $x_0=(n+1/2)vT_{br}$ where $n$ is an integer.
The period is measured as $T_{br}=0.98$ in the numerical evolution.
The density evolution of peak-peak collision between two breathing solitons is shown in Fig. \ref{pic-two} (a) when $n=7$.
Interestingly, two breathing solitons with larger breathing amplitude appear.
(As they are produced from original solitons by two collisions, they are called "after-two-collision breathing solitons".)
Also, it is found that two breathing solitons after dip-dip collision has very small breathing amplitude.
We compare the kinetic energy evolution in the cases of peak-peak and dip-dip collisions in Fig. \ref{pic-two} (b).
Compared with the initial breathing solitons, the derived ones have larger (or smaller) breathing amplitude after peak-peak (or dip-dip) collision.
Nevertheless, they have similar values of maximal kinetic energy, which coincides with the results shown in Fig. \ref{pic-phase} (b).
It indicates the velocity of breathing solitons after peak-peak collision is larger than the one after dip-dip collision.
Meanwhile, it is deduced that the collision at different parts of breathing solitons can produce solitons with different breathing dynamics.
Thus, we study the influence of position where they collide on the breathing amplitude of after-two-collision solitons, as shown in Fig. \ref{pic-two} (c).
The horizontal variable is $n_x=x_0/(vT_{br})$.
For an arbitrary integer $n$, when $n_x=n$ a peak-peak collision happens, and when $n_x=n+1/2$ a dip-dip collision happens.
From the plot, a dip-dip collision makes the breathing amplitude of after-two-collision solitons approach minimum, while a peak-peak collision makes it approach maximum.
Thus, one can use the peak-peak (or dip-dip) collision to generate solitons with a larger (or smaller) breathing amplitude.

\begin{figure}[htbp]
\centering
\includegraphics[width=86mm]{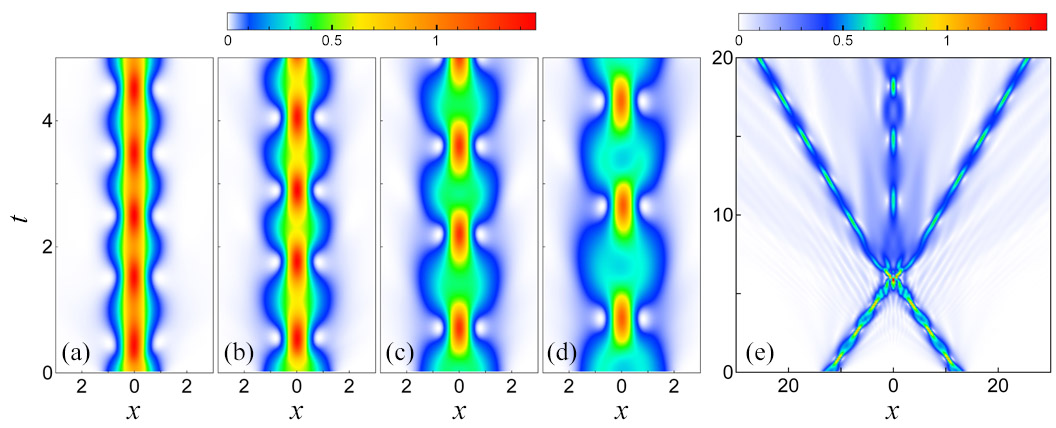}
\caption{(Color online) Density evolution plot of (a) after-one-collision soliton, (b) after-two-collision soliton, (c) after-three-collision soliton, (d) after-four-collision soliton, and (e) the peak-peak collision between two after-four-collision solitons.}
\label{pic-four}
\end{figure}

Similarly, one can go on the peak-peak collisions for more times to generate the solitons with a larger breathing amplitude.
The density evolutions of after-one-collision, after-two-collision, after-three-collision, and after-four-collision solitons are shown in Fig. \ref{pic-four} (a-d).
With the increasing number of collisions, the breathing behavior of solitons become more and more obvious, and their breathing period is longer and longer.
For the after-four-collision soliton in Fig. \ref{pic-four} (d), its dip has two maximums, which is different from other solitons.
When two after-four-collision solitons collide, its density evolution is shown in Fig. \ref{pic-four} (e), where three solitons appear after the collision.
Thus, a common after-five-collision soliton cannot be obtained in this way.
The appearance of splitting could be related to the two-maximum structure of after-four-collision soliton's dip, as the two maximums may split and then merge into the center soliton.
Among the derived three solitons, the center one has a breathing frequency changing over time, while the other ones breath stably.
We set the solitons on the sides as new initial solitons to make them collide at the peaks.
Two common solitons are obtained and the splitting phenomenon dose not appear.
Moreover, the robustness of breathing solitons is considered numerically.
The solitons shown in Figs. \ref{pic-four} (a-d) can keep stable under the noise with amplitude up to $50\%$, and the related result is not shown in this paper.

\section{Conclusion}

In summary, the breathing dynamics of dipolar solitons induced by their collision is studied numerically.
After the collision, there is a short-lived attracting process between solitons, which is induced by the long-range feature of dipolar effect.
This attraction makes the solitons momently wider and finally breathing.
The phase difference between two initial solitons has an obvious influence on the breathing amplitude, breathing frequency, and the transferred atom number of derived breathing solitons.
By considering the cases that two dipolar solitons move away from each other, we find that the initial phase difference has no influence on the difference of their breathing frequency.
Only when their initial atom number is different, their breathing frequency is different.
Thus, the difference of breathing frequency is directly induced by the difference of atom number, rather than the phase difference.
Also, the interaction between breathing solitons is studied.
The peak-peak collision between two breathing solitons generates solitons with larger breathing amplitude, while the dip-dip collision generates solitons with smaller one.
By the peak-peak collision, we successfully generate an after-four-collision breathing soliton.
However, three solitons instead of two ones appear after their collision, which indicates that an after-five-collision cannot be generated in the case we consider.
Our results pave a way to study the generation, interaction, and adjustment of breathing solitons in dipolar BECs and help to understand the influence of long-range feature of dipole-dipole interaction on solitons.

\section*{Acknowledgement}
This work was supported by National Natural Science Foundation of China 
(Contacts No. 11875220 and No. 12047502).

\end{document}